\documentclass[aps,prm, twocolumn,floatfix,amsmath,nofootinbib]{revtex4-1}

\usepackage{graphicx}
\usepackage[dvipsnames]{xcolor}
\usepackage{dsfont}
\usepackage[normalem]{ulem}
\usepackage{dcolumn}
\usepackage{multirow}
\usepackage{textcomp}
\usepackage{placeins}
\usepackage{subfigure}
\usepackage{float}
\hfuzz=\maxdimen
\newcommand{\vcn}[1]{\hat{\mathbf{#1}}}
\newcommand{\vcc}[1]{{\mathbf{#1}}}

\newcommand{\td}[1]{\mathrm{d} #1}

\providecommand{\abs}[1]{\lvert#1\rvert}

\providecommand{\ket}[1]{\ensuremath{\left\lvert \,#1\, \right\rangle}}

\providecommand{\braket}[2]{\ensuremath{\left\langle \,#1\, \rvert \,#2\, \right\rangle}}

\providecommand{\braketmat}[3]{\ensuremath{\left\langle #1 \right\lvert #2 \left\rvert #3 \right\rangle}}
\providecommand{\braketmatsmall}[3]{\ensuremath{\langle #1 \lvert #2 \rvert #3 \rangle}}

\newcommand{\eg}{e.g.,\@ }
\newcommand{\ie}{i.e.\@ }
\newcommand{\cf}{cf.\@ }
\newcommand{\etal}{\textit{et al.\@ }}
\newcommand{\abinitio}{\textit{ab initio}}

\newcommand{\erot}{\vcn{e}_\mathrm{rot}}
\newcommand{\eFM}{\vcn{e}_0}
\newcommand{\rr}{{\vcc{r}}}
\newcommand{\q}{{\vcc{q}}}
\newcommand{\knu}{{\vcc{k} \nu}}
\newcommand{\kn}{{\vcc{k} n}}
\newcommand{\Hso}{{\mathcal{H}_\mathrm{so}}}

\newcommand{\VolfacBxc}{\frac{B_\mathrm{xc}}{\Omega}}
\newcommand{\Volfac}{\frac{1}{\Omega}}
\newcommand{\Vol}{\Omega}

\begin{document}
\title{Electric dipole moment as descriptor for interfacial Dzyaloshinskii-Moriya interaction}
\author{Hongying Jia}
\email[Corresponding author: ]{h.jia@fz-juelich.de} 
\author{Bernd Zimmermann}
\email[Corresponding author: ]{be.zimmermann@fz-juelich.de} 
\author{Gregor Michalicek}
\author{Gustav Bihlmayer}
\author{Stefan Bl\"ugel}
\affiliation{Peter Gr\"unberg Institut and Institute for Advanced Simulation, Forschungszentrum J\"ulich and JARA, 52425 J\"ulich, Germany}


\begin{abstract}
Chiral magnets are of fundamental interest and have important technological ramifications. The origin of chiral magnets lies in the Dzyaloshinskii-Moriya interaction (DMI), an interaction whose experimental and theoretical determination is laborious. We derive an expression that identifies the electric dipole moment as descriptor for the systematic design of chiral magnetic multilayers. Using density functional theory calculations, we determine the DMI of (111)-oriented metallic ferromagnetic $Z$/Co/Pt multilayers of ultrathin films. The non-magnetic layer $Z$ determines the DMI at the Co-Pt interface. The results validate the electric and magnetic dipole moments as excellent descriptors. We found a linear relation between the electric dipole moment of Pt, the Allen electronegativity of $Z$, and the contribution of Pt to the total DMI.

\end{abstract}

\maketitle

\section{Introduction}

The Dzyaloshinskii-Moriya interaction (DMI)~\cite{Dzyaloshinsky:PCS:1958,Moriya:PR:1960} is the origin of chiral magnetism, a modern and active field of magnetism today. It is responsible for novel static and dynamical magnetic properties. In particular it drives the formation of chiral domain walls~\cite{Heide:PRB:2008,Thiaville:EPL:2012,Ryu:NN:2013} and skyrmions~\cite{Parkin:Science:2008,Heinze:NP:2011,Fert:NN:2013,Jiang:Science:2015,Boulle:NN:2016,Luchaire:NN:2016,Romming:PRL:2018,Herve:NC:2018}, which hold promise for applications in future information storage, data processing and neuromorphic devices. It generally promotes the stabilization of chiral non-collinear spin-textures.

A particularly important material class for applications is metallic films and multilayers of magnetic and heavy transition metals~\cite{Fert:MSF:1990,Zvyagin:JPCM:1991,Grigoriev:PRL:2008,Udvardi:PRL:2009,Hrabec:PRB:2014,Meyer:PRB:2017,Hoffmann:NC:2017,Bode:Nature:2007,Dupe:NC:2014,Dupe:NC:2016,Nandy:PRL:2016,Yagil:APL:2018,Zimmermann:PRB:2014}, as they are compatible with current manufacturing processes in spintronic devices. A particular example is the experimentally vividly pursued (111) oriented Co/Pt-based materials~\cite{Soumyanarayanan:NM:2017,Yang:PRL:2015,Grab:PRB:2018,Perini:PRB:2018,Shepley:PRB:2018,cao:Nanoscale:2018,Kim:NC:2018,Maccariello:NN:2018,Ajejas:APL:2017}. The huge combinatorics of different multilayers arising from the chemical composition, layer thicknesses, stacking sequences or growth conditions, to name a few, enables a detailed tuning of magnetic parameters, such as the DMI, which allows a flexible design of multilayers with very specific properties.

Qualitative insights into the formation of the DMI have been already gained by Moriya~\cite{Moriya:PR:1960}, or recently on the level of a tight-binding model~\cite{Kashid:PRB:2014}, but a simple model to predict the interfacial DMI quantitatively is currently unknown. Instead, the community relies either on experiments or \abinitio\ calculations to determine the DMI of a specific system. In both approaches, the procedure is rather involved and time-consuming, \eg Brillouin light scattering measurements need good statistics, or computing time intensive \abinitio\ calculations of typically rather large length-scale non-collinear magnetic structures need to be performed. In light of these restrictions, a systematic investigation of the large combinatorial space of multilayers is currently unthinkable.

The quest is open for simple descriptors which are faster to measure or calculate, but yet allow a reliable estimation of the DMI. Their application would allow for a high-throughput screening of materials and narrow the search-space considerably. Some quantities have already been proposed as descriptors for the DMI, mostly focusing on the magnetic layer:
Belabbes~\etal~\cite{Belabbes:PRL:2016} report on a correlation between the DMI and the spin magnetic moments of $3d$ atoms in $3d$ ultrathin films on $5d$ substrates. Kim~\etal~\cite{Kim:NC:2018} conclude on a correlation between the DMI and many other magnetic properties from temperature-dependent studies on the same material, \eg emphasizing on the anisotropy of the orbital magnetic moment and the magnetic dipole moment of the ferromagnetic metal in experiment. Modifying the interface dipole or Rashba fields through charge transfer by oxidization of the magnetic layer was proposed as a way to alter the DMI~\cite{Belabbes:SR:2016,Chaves:PRB:2019}. Shifting the focus to the heavy-metal substrate, Ryu and coworkers~\cite{Ryu:NN:2013,Ryu:NC:2014} argue that the DMI is closely tied to the induced magnetic moment at the $5d$ atom, whereas other studies report that such a direct relation could not be confirmed~\cite{Rowan:SR:2017, Yang:PRL:2015, Sandratskii:PRB:2017, Jia:PRB:2018}. Experimental observation of the correlation between the DMI and work function of non-magnetic layers in metallic magnetic trilayers is reported by Park~\etal \cite{Park:NAM:2018}.

In this paper, we explore possible descriptors, in particular the local electric and intra-atomic magnetic dipole moments. By an analytic derivation in perturbation theory, we identify a link, to leading order, between the DMI and the electric dipole moment in the sense that both quantities emerge from the same electronic states. We confirm our finding by first-principles calculations based on density functional theory (DFT) on (111) oriented magnetic multilayers (MMLs) composed of $Z$/Co/Pt, where the layer $Z$ is one of the $4d$ transition metals (Y--Pd), the noble metals (Cu, Ag and Au), or one of the post-transition metals Zn and Cd. This is a suitable test-set, because, as we show below, the DMI is modified drastically as function of $Z$ to positive and negative values. We find the largest correlation between the DMI and the electric dipole moment of either Co or Pt, and the sign is predicted correctly in 12 out of 13 different MMLs. To a lesser degree a correlation between the DMI and the (induced) magnetic dipole moment on Pt is found. We do not find a correlation between induced spin- or orbital moments at Pt.

\section{Theoretical relations between DMI and dipole moments}
\label{sec:theoretical_derivation}

The micromagnetic energy functional describing the DMI for (111)-oriented MMLs in terms of a continuous magnetization field $\mathbf{m}(\mathbf{r})$ is given as
\begin{equation}
E_\mathrm{DM}[\mathbf{m}] =\int \mathrm{d}^2 r \, D_\mathrm{s} \left[ \mathbf{m} (\nabla \cdot \mathbf{m}) -   (\mathbf{m} \cdot \nabla) \mathbf{m} \right]_z\, ,
\label{eq:1}
\end{equation}
and is usually termed the \emph{interfacial DMI} with an interface DMI constant $D_\mathrm{s}$.\footnote{The subscript $s$, derived from \emph{surface}, denotes the interfacial DMI constant. Typical values for $D_\mathrm{s}$ are on the order of a few pJ/m for metallic interfaces, and a conversion to a DMI acting on a volume of magnetic material reads $D = D_\mathrm{s}/t$, where $t$ is the thickness of the magnetic volume.} The latter relates to the change of spin-orbit energy upon twisting of the magnetization. It can be determined from a microscopic model, \eg a DFT model, by the $\q$-linear part of the energy change if a spiraling magnetic texture of wave-vector $\q||\hat{x}$ and rotation axis $\erot=\hat{y}$ is imposed [$\hat{z}$ is the out-of-plane direction; see Fig.~\ref{fig:DMI+properties}(b)]~\cite{Schweflinghaus:PRB:2016},
\begin{equation}
    D_\mathrm{s} = \Volfac \left. \frac{\partial E^\mathrm{DFT}_\mathrm{DM}(q\,\hat{x},\hat{y})}{\partial q} \right\rvert_{q=0}\, , \label{eq:model:DMI_q_derivative:main}
\end{equation}
with $\Vol$ the interface area per unit cell normal to the MML. We treat non-collinearity and spin-orbit coupling in perturbation theory and arrive at (Appendix~\ref{sec:appendix:Modelderivation})

\begin{equation}\!
    D_\mathrm{s} = \Volfac \sum_{\knu}^\mathrm{occ.}\!\! \sum_{\nu'}^\mathrm{all} \frac{\braketmat{\psi^0_{\knu}}{\Hso}{\psi^0_{\knu'}} \braketmat{\psi^0_{\knu'}}{\mathcal{T}_y  x}{\psi^0_\knu} }{\epsilon^0_{\knu} - \epsilon^0_{\knu'}}\! +\! \text{c.c.},\label{eq:model:Ds:perturb1:main}
\end{equation}
where $\epsilon^0_{\vcc{k}\nu}$ and $\psi^0_{\knu}$ are the (unperturbed) band energy and wavefunction, respectively, of state $\nu$ at crystal momentum $\vcc{k}$ of the ferromagnetic state, $\Hso$ is the spin-orbit Hamiltonian, $\boldsymbol{\mathcal{T}} = -\boldsymbol{\sigma} \times \vcc{B}_\mathrm{xc}^{0}$ is the torque operator, and $\vcc{B}_\mathrm{xc}$ is the exchange field~\footnote{A similar equation has been derived by Freimuth \etal~\cite{Freimuth:CondMat:2014}, where SOC was not treated in perturbation theory but entered through the wave-functions.}. The summations are performed over occupied states (occ.) as well as occupied and unoccupied (all) states.

The nominator is a product of spin-orbit and spin-torque-moment matrix elements, respectively. They are non-vanishing only if certain relations between the states $\ket{\psi^0_{\knu}}$ and $\ket{\psi^0_{\knu'}}$ are fulfilled (selection rules), which we discuss next. In what follows, we separate the spatial and spin-parts, $\ket{\psi_{\nu}} = \ket{\varphi_n} \otimes \ket{\sigma}$, and drop the crystal momentum $\vcc{k}$ and superscript ``$0$'' to simplify the notation. $\nu = (n,\sigma)$ is a multi-index containing the band index $n$ and spin index $\sigma$. 

For the nominator the following hold: (i) The spin-torque-moment contains the Pauli-matrix $\sigma_x$ which selects spin-flip contributions, \ie the two states must be of different spin-character.  An example that satisfies this condition is $\ket{\psi_\nu} = \ket{\varphi_1} \otimes \ket{\uparrow}$ and $\ket{\psi_{\nu'}} = \ket{\varphi_2} \otimes \ket{\downarrow}$. Assuming in addition that the exchange field is constant in space, the nominator in Eq.~\eqref{eq:model:Ds:perturb1:main} simplifies to $B_\mathrm{xc}\braketmat{\varphi_1}{\mathcal{H}_\mathrm{so}^{\uparrow \downarrow}}{\varphi_2}\braketmat{\varphi_2}{x}{\varphi_1}$. (ii) The spin-flip part of spin-orbit coupling, $\mathcal{H}_\mathrm{so}^{\uparrow \downarrow}$, selects transitions between states where the angular momentum index does not change, $|\ell-\ell'|=0$. (iii) The $x$ operator selects transitions where the angular momentum index changes by 1, $|\ell-\ell'|=1$. The latter two conditions are mutually excluding, and an overall non-vanishing DMI is only possible if at least one of the states is a mixed state of two orbital characters, say $\ket{\varphi_1} = \alpha \ket{s} + \beta \ket{p_z}$ and $\ket{\varphi_2} = \ket{p_x}$, so that the above nominator turns into $\alpha \beta^*\,B_\mathrm{xc}\, \braketmat{p_z}{\mathcal{H}_\mathrm{so}^{\uparrow \downarrow}}{p_x} \, \braketmat{p_x}{x}{s} \neq 0$.

In the next step, we analyze all possible combinations of $\ket{\varphi_1}$ and $\ket{\varphi_2}$ assuming a basis of $s$, $p$, and $d$ orbitals, which contribute to the DMI, exploiting one basic symmetry operation of (111) oriented multilayers with $C_{3v}$ symmetry, namely a mirror plane that we choose perpendicular to $x$. As a consequence, the wavefunctions of the system are either even $(+)$ or odd $(-)$ under this symmetry operation, $\psi(-x,y,z) = \pm \psi(x,y,z)$. Accordingly we classify the atomic orbitals to be either of even ($s$, $p_y$, $p_z$, $d_{z^2}$, $d_{yz}$, $d_{x^2-y^2}$) or odd ($p_x$, $d_{xz}$, $d_{xy}$) symmetry, and hence only superpositions among even or odd states, respectively, are allowed candidates for $\ket{\varphi_1}$. In Table~\ref{tab:transitions:DMI} all possible transitions that yield a finite contribution to the DMI are summarized.

\begin{table}
\begin{ruledtabular}
    \centering
    \caption{Possible combinations of states $\ket{\varphi_1}$ and $\ket{\varphi_2}$ that yield a finite contribution to the DMI; see Eq.~\eqref{eq:model:Ds:perturb1:main} and text for details. We denote the states that also contribute to a finite electric dipole moment, $p^\mathrm{el}_z$, by an asterisk.}
    \label{tab:transitions:DMI}
    \begin{tabular}{rclc|c}
\multicolumn{4}{c|}{$\ket{\varphi_1}$}            & $\ket{\varphi_2}$  \\[1.1ex] \hline
$\alpha \ket{s}   $&$+$&$ \beta \ket{p_z}$        & (*) & $\ket{p_x}$ \\
$\alpha \ket{p_y} $&$+$&$ \beta \ket{d_{yz}}$     & (*) & $\ket{d_{xy}}$ \\
$\alpha \ket{p_z} $&$+$&$ \beta \ket{d_{z^2}}$    & (*) & $\ket{p_x}$ or $\ket{d_{xz}}$ \\
$\alpha \ket{p_z} $&$+$&$ \beta \ket{d_{x^2-y^2}}$&     & $\ket{p_x}$ or $\ket{d_{xz}}$ \\
$\alpha \ket{p_x} $&$+$&$ \beta \ket{d_{xz}}$     & (*) & $\ket{p_z}$ or $\ket{d_{x^2-y^2}}$ or $\ket{d_{z^2}}$ \\
    \end{tabular}
    \end{ruledtabular}
\end{table}

It is interesting to note that if we additionally had a mirror plane perpendicular to $z$ present in the system, all superpositions for $\ket{\varphi_1}$ in Table~\ref{tab:transitions:DMI} would be prohibited by symmetry and the DMI would vanish. This corresponds to Moriya's symmetry rules~\cite{Moriya:PR:1960}.

Our analysis shows that hybridization is the key to obtaining a finite DMI. Equation~\eqref{eq:model:Ds:perturb1:main} represents an important result and provides physical insights into the mechanism of formation of the DMI. For example, the appearance of the position operator calls for a relation to the electric dipole moment 
\begin{equation}
    \vcc{p}^\mathrm{el} = -e \sum_{\knu} \braketmat{\psi_\knu}{\vcc{r}}{\psi_\knu} = \int_{\Omega_\mathrm{MT}}{ \rho(\vcc{r})\,\vcc{r}}\, \mathrm{d}\mathbf{r}\, , \label{electricDipoleDefEq}
\end{equation}
of which only the out-of-plane direction does not vanish due to the symmetry of the system. 
$e$ is the (positive) elementary charge, $\rho(\mathbf{r})$ is the electron charge density, the integration is performed over the volume of a sphere, typically the muffin-tin (MT) sphere $\Omega_\mathrm{MT}$, around the considered atom and the real-space vector $\vcc{r}$ is measured with respect to the center of this sphere. Similarly to the arguments presented above, the states $\psi_\nu$ that contribute must be a superposition of two states with $\Delta \ell = 1$. Evaluation of these matrix elements reveals that four of the five possible superpositions $\varphi_1$ from Table~\ref{tab:transitions:DMI} constitute the finite $p_z^\mathrm{el}$. If we were able to change the superpositions $\varphi_1$ continuously by some external means, both, the DMI and $p_z^\mathrm{el}$ would scale similarly, \eg to leading order with $\alpha \beta^*$.

Expressing the spin-torque-operator in \eqref{eq:model:Ds:perturb1:main} beyond the constant exchange-correlation field approximation by including in addition the spherically symmetric contribution around the atoms, $\vcc{B}_{xc}(r)$, we obtain qualitatively the same result as presented above (Appendix~\ref{sec:appendix:Bxcassumptions}).

\section{Computational details and Method}
\label{sec:computmethods}

The experimental bulk lattice constant of Pt ($392~\mathrm{pm}$) corresponding to an (111) in-plane lattice constant of $a=277~\mathrm{pm}$ was chosen and structural optimizations of all interlayer distances were performed in scalar-relativistic approximation using a mixed LDA/GGA spin-density-functional~\cite{Santis:PRB:2007} (LDA and GGA stand for local density approximation and generalized gradient approximation, respectively). Subsequent calculations of the electronic and magnetic properties employed the LDA~\cite{Vosko:JP:1980}. We converged the charge density and spin density in a scalar-relativistic approximation, sampling the full Brillouin zone (BZ) by $(24\times24\times10)$ $\mathbf{k}$-points if not stated otherwise.
For the extraction of the DMI, we first performed self-consistent calculations of homogeneous spin-spirals with a wave-vector $\mathbf{q}$ along the $\Gamma$--$M$ high-symmetry line of the BZ and $\vert \mathbf{q} \vert \leq 0.1 \times 2\pi/a$. The DMI was determined by including SOC in first-order perturbation theory on top of a scalar-relativistic spin-spiral calculation~\cite{Heide:Physica:2009} using ($48\times48\times20$) $\mathbf{k}$-points (see Refs.~\cite{Zimmermann:PRB:2014,Jia:PRB:2018} for details). The calculation of the magnetic dipole moment was performed on a ($35\times35\times18$) $\mathbf{k}$-point set which included the $\Gamma$ point, and SOC was either neglected or included in the self-consistent calculations with collinear magnetization along $\hat{z}$. In all calculations, the 4$s$ and 4$p$ states of Y, as well as the 4$p$ states of Zr, Nb, and Mo were represented by extending the conventional LAPW basis set with local orbitals~\cite{Singh:PRB:1991,Michalicek:CPC:2013}.

\section{Results}
\subsection{Correlation to electric dipoles}

\begin{figure}[!t]
\centering
\includegraphics[width=0.45\textwidth]{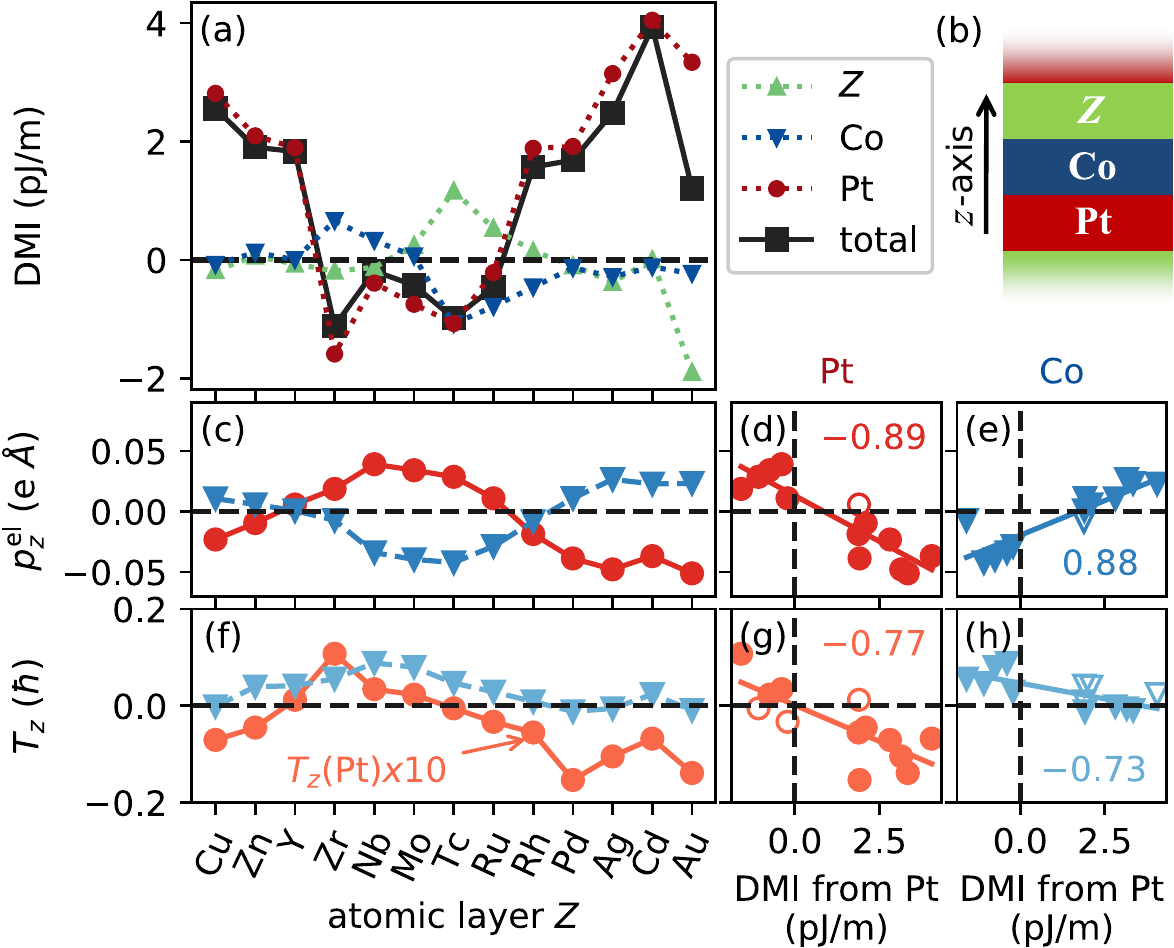}
\caption{Properties of magnetic multilayers $Z$/Co/Pt for various chemical elements $Z$. (a) The total DMI, and broken down into spin-orbit contributions of the different layers. (b) Sketch of the MMLs. (c-e) Electric dipole moment $p^\mathrm{el}_z$ and (f-h) magnetic dipole moment $T_z$ against $Z$ and the DMI. $T_z$ of Pt has been multiplied by a factor 10 for better visibility. Lines in (a), (c), and (f) are guides to the eye. Solid lines in (d,e,g,h) indicate least-squares fits, and the Pearson correlation coefficient is displayed in the panels. Data where the sign between the DMI and the $p^\mathrm{el}_z$, $T_z$ is wrong are indicated by open symbols. See Supplemental Material for the data presented.}
\label{fig:DMI+properties}
\end{figure}

In order to test our derivation of the correlation between $p_z$ and $D_\mathrm{s}$ on a realistic test-set, we perform density-functional theory (DFT) calculations using the full-potential linearized augmented plane wave (FLAPW) method, as implemented in the FLEUR code~\cite{FLEUR}. The unit cell of our magnetic multilayer consists of three monolayers, namely Co sandwiched between Pt and $Z$ [see Fig.~\ref{fig:DMI+properties}(b)], where $Z$ is a $4d$ transition metal (Y--Pd), a noble metal (Cu, Ag, and Au), or one of the post-transition metals Zn or Cd. We assume a fcc stacking of the layers and a ferromagnetic order of all Co moments for a better comparability of our results, although also a synthetic antiferromagnetic coupling between adjacent Co layers might be energetically favorable for some of the here studied multilayers. See Sec.~\ref{sec:computmethods} for computational details.

From the analysis of the DMI as computed \abinitio, we deduce a strong dependence of the magnitude and even the sign of the total DMI on the chemical element of the third atomic layer, $Z$ [see Fig.~\ref{fig:DMI+properties}(a)]. In particular, within the 4$d$ series, the modification of the DMI becomes evident: we obtain small negative values when $Z$ is an early transition metal ($D_\mathrm{s}=-1.11$~pJ/m for $Z=\mathrm{Zr}$), followed by a rather continuous change towards positive and large values when the $d$ shell gets filled ($D_\mathrm{s}=3.93$~pJ/m for Cd). We break up the total DMI, $D_\mathrm{s}$, of the system into contributions from the three atomic layers, denoted as $D_\mathrm{s}^{Z}$, $D_\mathrm{s}^\text{Co}$, and $D_\mathrm{s}^\text{Pt}$ for $Z$, Co, and Pt, respectively, by switching spin-orbit coupling on only in a single layer at a time. Within first-order perturbation theory in spin-orbit coupling (SOC), which we apply here, this decomposition is exact. The DMI originating from SOC of the Pt layer constitutes the dominant contribution [\cf Fig.~\ref{fig:DMI+properties}(a)] due to its large atomic number. Remarkably, this contribution changes drastically as the chemical type of the third layer is varied and, in effect, determines the overall trend of the DMI of the entire stack. This is surprising because it is commonly believed that the DMI at the Co/Pt interface can merely be changed by external means, \eg by controlling the interface quality~\cite{Lavrijsen:PRB:2015,Zimmermann:APL:2018}. In contrast, the Co layer and third layer $Z$ only contribute little to the DMI. Au/Co/Pt represents an exception, because Au gives a rather large contribution ($-1.90$~pJ/m), but of opposite sign as compared to Pt ($3.34$~pJ/m), hence reducing the total DMI by more than a factor 2.

Overall, our results cannot be explained by a linear additive superposition of a constant DMI from the Pt/Co interface and a varying DMI from the Co/$Z$ interface, as most of the changes are originating from the former. It is not surprising that the separation into two individual interfaces breaks down for individual layers that are very thin, \ie one atomic layer in the present case. Interestingly, the DMI from Pt can be even enhanced (\eg by approximately 40\% in Pt/Co/Cd) as compared to the DMI of a single Co/Pt interface ($D_\mathrm{s} \approx 2.8$~pJ/m~\cite{Freimuth:CondMat:2014}).

Next we investigate how this drastic change of the DMI between Co and Pt correlates with our predicted descriptor, the electric dipole moment. In Fig.~\ref{fig:DMI+properties}(c), we present the \abinitio\ computed local electric dipole moments in the spheres of Co and Pt, see \eqref{electricDipoleDefEq}. They are nearly of the same magnitude and opposite sign and exhibit characteristic sign-changes around Y and Rh, similarly to the DMI. Indeed, as Figs.~\ref{fig:DMI+properties}(d,e) show, there exists a linear relationship between the DMI and $p^\mathrm{el}_z$ of Co and Pt, respectively. The overall correlation between the DMI and $p^\mathrm{el}_z$ is very large, as expressed by the Pearson's coefficient $\vert R \vert=0.89$ and 0.88, respectively [see Figs.~\ref{fig:DMI+properties}(d,e)]. Also the sign of the DMI correlates with the sign of $p^\mathrm{el}_z$ in all cases except one (Y/Co/Pt). By means of a least-squares fit we find
\begin{eqnarray}
    D_\mathrm{s}^\text{Pt} &\approx& \left(-0.53~\frac{\text{J}}{e~\text{m}^2} \right) ~ p^\mathrm{el}_z(\text{Pt})  + 0.96~\frac{\text{pJ}}{\text{m}}\\
    &\approx& \left(0.67~\frac{\text{J}}{e~\text{m}^2} \right) ~ p^\mathrm{el}_z(\text{Co})  + 1.6~\frac{\text{pJ}}{\text{m}}~,
\end{eqnarray}
where the electric dipole moment is given in units of $e~\mathrm{m}$.

\subsection{Relation to magnetic dipoles}

Returning to our analysis of \eqref{eq:model:Ds:perturb1:main} and considering a non-spherical contribution to $\vcc{B}_\mathrm{xc}(\vcc{r})$ yields a correction term for the spin-torque moment of the form $\delta \mathcal{T}_y x \sim Q_{zx} \sigma_x$ (Appendix~\ref{sec:appendix:Bxcassumptions}) and can be identified as a contribution to the magnetic dipole moment \cite{Stohr:JMMM:1999},
\begin{equation}
  \vcc{T} = \frac{\hbar}{2} \sum_{\knu} \braketmat{\psi_\knu}{\underline{\underline{Q}} \cdot \boldsymbol{\sigma}}{\psi_\knu}, \label{eq:Tz:QdotS}
\end{equation}
which should thus contribute to the DMI. Here, $Q_{ij} = \delta_{ij} - 3 \hat{r}_i \hat{r}_j$, $i,j\in\{x,y,z\}$, are the components of the (dimensionless) quadrupole tensor and $\boldsymbol{\sigma} = \left( \sigma_x, \sigma_y, \sigma_z \right)^\mathrm{T}$ is the vector of Pauli matrices. The magnetic dipole $\vcc{T}$ reflects the asphericity of the magnetization density in the muffin-tin sphere around an atom, and has two contributions: one is induced by the crystal field and the other by SOC~\cite{Suzuki:PhysLettA:2019}. Our correction term $Q_{zx} \sigma_x$ is related to the latter.

We computed $T_z$ for all multilayers with and without SOC and find that the SOC induced changes in $T_z$ are on the order of 0.001~$\hbar$, which is 1--2 orders of magnitude smaller than the crystal field part, even for Pt. Moreover, we do not find a strong correlation to the DMI, which renders the correction described above unimportant. Interestingly, we find a sizable correlation between the crystal-field induced $T_z$ and the DMI [$\vert R \vert=0.77$ and 0.73 for Pt and Co, respectively; see Figs.~\ref{fig:DMI+properties}(f-h)], with the caveat that $T_z$ of Co is not able to predict the sign of the DMI with high fidelity: There is an offset in the data, which is related to the fact that a free-standing Co monolayer exhibits a finite $T_z$~\cite{Oguchi:PRB:2004} due to the strong asphericity in the crystal field (\ie $x$ and $z$ directions are inequivalent), but the DMI vanishes due to the presence of structure inversion symmetry ($+z$ and $-z$ directions are equivalent). Instead, we find a better correlation with $T_z$ of Pt, which is an induced magnetic dipole moment and the structure-inversion asymmetry is implicitly imprinted in its existence in this case. However, this sizable correlation between the crystal-field part of $T_z$ and the DMI cannot be explained by Eq.~\eqref{eq:model:Ds:perturb1:main}, and it might be necessary to develop a non-local theory or, since the position operator $\mathbf{r}$ is not a proper operator in the Hilbert space of periodic solids, turn to the corresponding Berry-phase expressions~\cite{Freimuth:CondMat:2014}, which is beyond the scope of this paper.

\subsection{Relation to magnetic moments}

\begin{figure}[t]
\centering
\includegraphics[width=85mm]{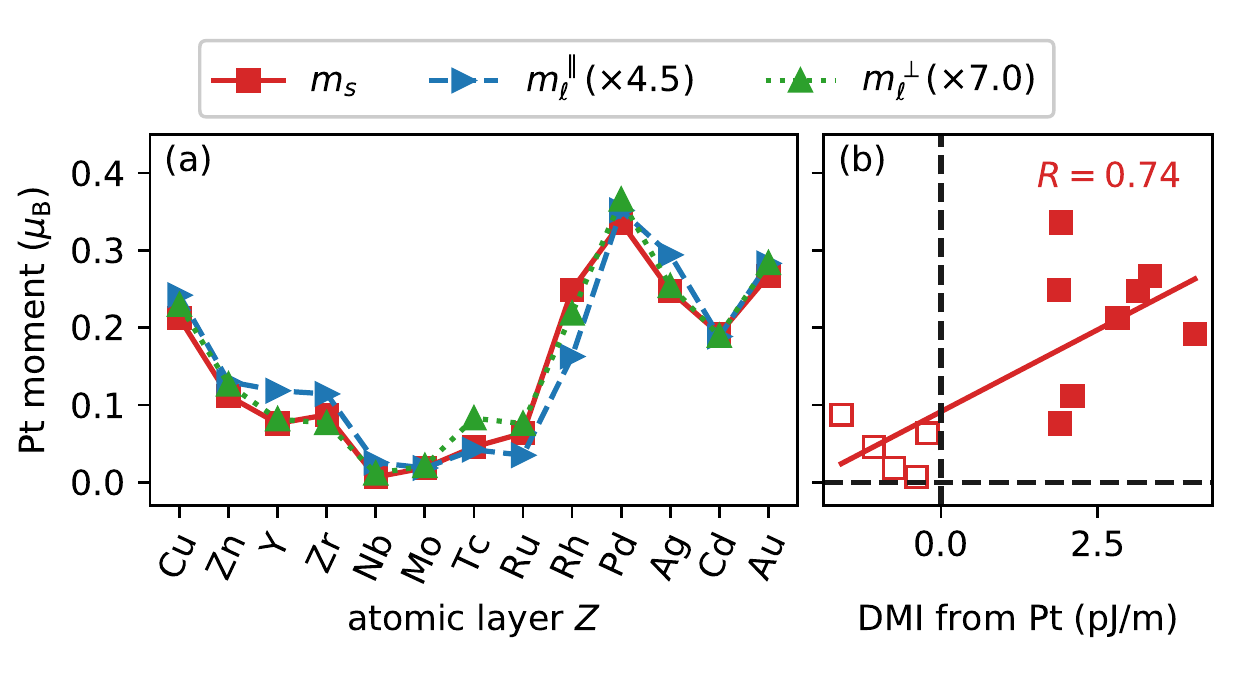}
\caption{(a) Induced Pt spin-moments ($m_s$), orbital moments for in-plane ($m_\mathrm{\ell}^\parallel$) and out-of-plane ($m_\mathrm{\ell}^\perp$) magnetization directions in $Z$/Co/Pt MMLs. (b) Induced Pt spin moments against the DMI with least-squares fit and Pearson's correlation coefficient $R$. A wrong sign between the DMI and $m_s$ is indicated by open symbols. See Supplemental Material for the data presented.}
\label{fig:magmom}
\end{figure}

To shed light on the controversial debate on the relationship between the induced spin moment of Pt and the DMI, we investigate the magnetic spin and orbital moments of Pt. All moments are positive, meaning they are parallel to the Co moments, and proportional to each other with $m_\mathrm{\ell}^\parallel/m_s = 0.14$ and $m_\mathrm{\ell}^\perp/m_s = 0.22$, where the magnetization lies in-plane or along the out-of-plane direction, respectively (see Fig.~\ref{fig:magmom}). The correlation coefficient ($R=0.74$) between the DMI and $m_s$ is rather high, but we cannot deduce a causal relationship between these two quantities. Our reasoning is underpinned by the fact that the sign of the DMI cannot be correctly reproduced by the magnetic moments in five out of 13 cases. Our results also highlight that neither the orbital moments, nor an anisotropy of the orbital moments is correlated to the DMI. In addition, we also do not find a sizable correlation between DMI and the electronic charge of Co or Pt.

\section{Discussion and Conclusion}

We calculated the DMI of periodic $Z$/Co/Pt(111) multilayers of monatomic layer thickness and compare our results with the experimental investigation of $Z$/Co/Pt(111) magnetic trilayers with film thicknesses between 11 (Pt) and 5 (Co) atomic layers reported by Park \etal~\cite{Park:NAM:2018}. We compare those systems which have a ferromagnetic ground state in both our and their work: the total DMI is 1.42 (2.56) pJ/m, 0.44 (1.68) pJ/m, 0.26 (1.21) pJ/m for Cu/Co/Pt, Pd/Co/Pt, and Au/Co/Pt, in Ref.~\cite{Park:NAM:2018} (in this work). First, we note that the signs of DMI in our theoretical results are the same as in the experiment. Second, the relative values in the two papers are also consistent ($D_\mathrm{Cu}>D_\mathrm{Pd}>D_\mathrm{Au}$). In our work, however, the overall magnitude is up to a factor 5 higher than in the experiment. Since our values do not appear unreasonably high when compared to the DMI of a single Co/Pt interface with ultrathin Co layers  ($D_\text{s} = 1.7$~pJ/m)~\cite{Belmeguenai:PRB:2015}, we conjecture that the major difference arises from the different layer thicknesses and the related differences in the lattice strain. Additional factors responsible for the difference include our focus on atomically sharp trilayers, while in experimentally produced multilayers the vertical texture will certainly differ, \eg due to slight intermixing effects at the interfaces~\cite{Zimmermann:APL:2018} or a different stacking sequence. Another reason is that we have treated periodic multilayers where the non-magnetic spacer layer hybridizes with both Co and Pt layers, but in Ref.~\cite{Park:NAM:2018} trilayer systems are treated where only the Co layer is affected by the spacer layer.
	
	\begin{figure}[!h]
		\centering
		\includegraphics[width=85mm]{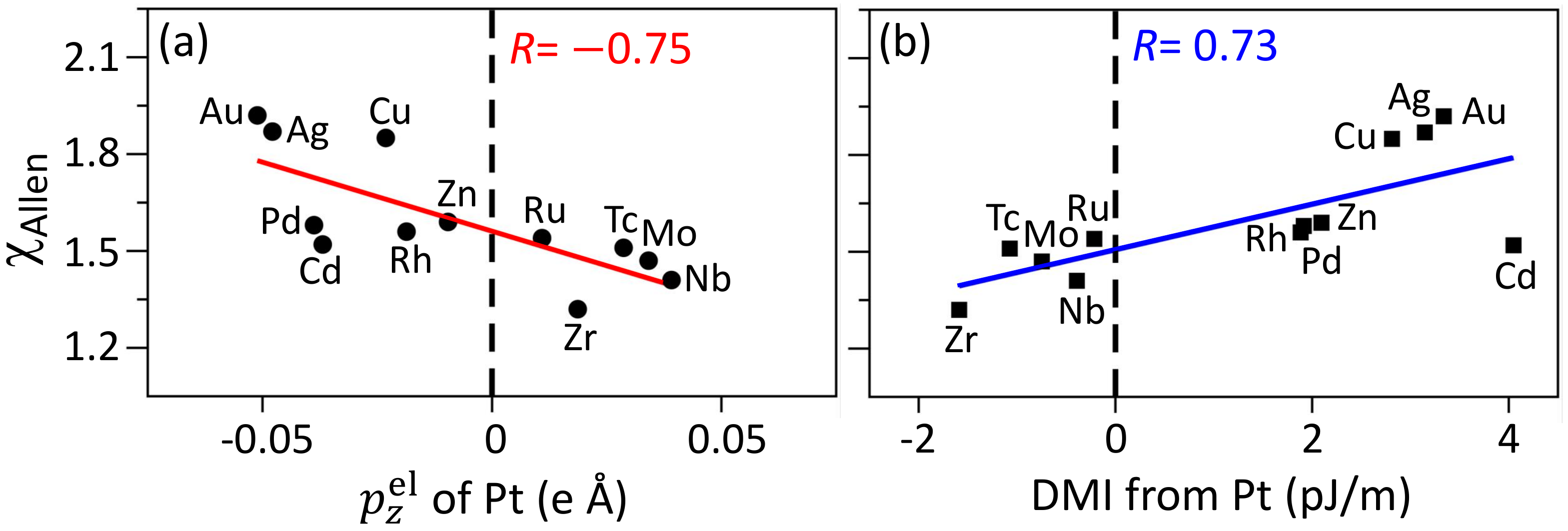}
		\caption{Correlation between Allen electronegativity in Pauling units and  (a) the electric dipole moment and (b) the DMI strength of Pt of magnetic multilayers $Z$/Co/Pt for various chemical elements $Z$. Not included is element Y. See Supplemental Material for the data presented.
			\label{fig:AllenEN_vs_DMI}}
		
	\end{figure}
	
	As a simplified guide to materials systems design we investigated the relation of established tabulated data effecting the electric dipole and thus the DMI of Pt. For example, we also investigated the relationship between the DMI and the  electrical dipole moment with respect to the workfunction changes between Co and $Z$ and between Pt and $Z$, but unlike Park~\etal~\cite{Park:NAM:2018} we could not find any correlation, at least not for our periodic multilayers of ultrathin films. Instead we found useful relations  (Pearson correlation coefficient $R> 0.7$), shown in Fig.~3, between the DMI as well as the  electric dipole moment of Pt and the tabulated Allen electronegativity~\cite{Allen:ACS:1989}, $\chi_\text{Allen}$, for $4d$ transition-elements $Z$ of the groups 4 to 10 of the periodic table, the group of noble metals (group 11) and to a certain extent also to the elements of the group 12 ($Z=$ Zn, and Cd). We found large deviations for Y, element of  group 3. From Fig.~3(b) we read-off  the relation
	\begin{equation}
	D_\mathrm{s}^\text{Pt} \approx \left(7.62~\frac{\text{pJ}}{\text{m}} \right) ~ \chi_\text{Allen}(Z)  - 10.89~\frac{\text{pJ}}{\text{m}}\\
	\end{equation}
	with parameters obtained by  means of a least-squares fit. The  electronegativity is entered in Pauling units. Since according to Fig.~1(a) the total DMI follows the DMI contribution of Pt with the exception of Au,  we accordingly find a predictive relation between the Allen electronegativity and the total DMI for the remaining systems. Similar results we obtained for the Mulliken electronegativity~\cite{Mulliken:1934:JCP}, while no suitable correlation was found for the Pauling electronegativity~\cite{Pauling:1932:ACS}.

	Our work points out a few important guidelines to realize the maximal impact on the DMI:\\
	(i) The layers adjacent to the heavy element (in our case: Pt) should have a large difference in the Allen  electronegativity to maximize the dipole $p_z^\mathrm{el}$ at this element.\\
	(ii) The heavy (Pt) layer should be thin with sharp interfaces in particular to the magnetic (Co) layer, but not too thin. There should be a compromise between the accumulation of DMI strength by increasing the number of Pt layers and the finite screening length beyond which the electrical dipole moment in Pt cannot be changed significantly by the second non-magnetic layer $Z$. The second non-magnetic element can consist of several atomic layers.
	
	For the external manipulation of the dipole with an electric field our findings suggest that the field should act mainly on the heavy atomic species. Because of the screening  of the field by the conductive electrons, however, the external manipulation  of the electrical dipole by an electrical field is usually limited to layer thicknesses of a few atomic layers. On the other hand, there are experiments carried out with trilayers where one layer is a simple oxide that acts as electrode, \eg Au/Fe/MgO~\cite{Nawaoka:APE:2015}. A change of DMI could be detected upon application of an electric field. Our findings suggest that an optimal control could be achieved by reverting the position of Au and Fe, \ie Fe/Au/MgO, to maximize the field acting on the Au rather than the Fe.

In conclusion, we derived an analytic expression for the Dzyaloshinskii-Moriya interaction (DMI) based on perturbation theory from the ferromagnetic state and postulated a relation to the electric dipole moment. Subsequent \abinitio\ calculations on magnetic multilayers of the type $Z$/Co/Pt indeed showed that the interfacial DMI, which takes values between $-1$ and 4~pJ/m, strongly correlates to the electric dipole moment $p^\mathrm{el}_z$. Since the electric dipole can be calculated rather quickly, we propose the evaluation of this quantity for a screening of chiral multilayer systems. In contrast, the intra-atomic magnetic dipole moment $T_z$ and induced spin-moments correlate less and are, in the latter case, not able to predict the sign-change of DMI within the $4d$ series of $Z$.

\section{Acknowledgments}

We thank Tatsuya Shishidou and Tamio Oguchi for fruitful discussions and computational validation of the magnetic dipole vector.  We thank Frank Freimuth, Jan-Philipp Hanke, and Yuriy Mokrousov for in-depth discussions on the DMI interaction. We gratefully acknowledge financial support from the DARPA TEE program through grant MIPR (No. HR0011831554) from DOI, from the European Union H2020-INFRAEDI-2018-1 program (Grant No. 824143, project “MaX - Materials at the Exascale”), from Deutsche  For\-schungs\-gemeinschaft (DFG) through SPP 2137 ``Skyrmionics" (Project BL 444/16), from the Collaborative Research Centers SFB 1238 (Project C01) as well as computing resources at the supercomputers JURECA at J\"{u}lich Supercomputing Centre and JARA-HPC from RWTH Aachen University. 

\appendix
\section{DMI in perturbation theory}
\label{sec:appendix:Modelderivation}

Our aim is to derive \eqref{eq:model:Ds:perturb1:main} from the main text starting with the well-established relation between the interface DMI constant, $D_\mathrm{s}$, and the change of the total energy $E^\mathrm{DFT}_\text{DM}$~\cite{Schweflinghaus:PRB:2016}, \eqref{eq:model:DMI_q_derivative:main} from the main text, which we repeat here for convenience:
\begin{equation}
    D_\mathrm{s} = \Volfac \left. \frac{\partial E^\mathrm{DFT}_\mathrm{DM}(q\,\hat{x},\hat{y})}{\partial q} \right\rvert_{q=0}\, . \label{eq:model:DMI_q_derivative:app}
\end{equation}
The latter is the SOC contribution to the total energy relative to the collinear (ferromagnetic) state for a spin-spiral state of a general wavevector $\q$ and rotation axis $\erot$, which is given in first-order perturbation theory as
\begin{eqnarray}
    E_\mathrm{DM}(\q,\erot) = \sum_{\vcc{k}, \nu}^\mathrm{occ.}{\braketmat{\psi_{\knu}(\q)}{\Hso}{\psi_{\knu}(\q)}}\, , 
\label{eq:model:EDM}
\end{eqnarray}
where $\psi_{\knu}$ is the unperturbed wavefunction of state $\nu$ at crystal momentum $\vcc{k}$, $\Hso$ is the spin-orbit Hamiltonian, and the summation is performed over occupied states. Since we are interested in the energy difference due to the infinitesimal deviation of the magnetization from the collinear state by introducing  a long-wavelength spin-spiral,  we express the wavefunctions $\psi_{\knu}(\q)$ of the non-collinear state in terms of the ferromagnetic state $\psi^0_{\knu}$  in first-order perturbation theory,

\begin{eqnarray}
    \psi_{\knu}(\q,\rr) &=& \psi^0_{\knu}(\rr) + \label{eq:model:psi-1st-order} \\
    && \sum_{\vcc{k}'\nu' (\neq \vcc{k}\nu)}{\!\!\! \frac{\braketmat{\psi^0_{\vcc{k}'\nu'}}{\mathcal{O}_\mathrm{xc}(\q)}{\psi^0_\knu}}{\epsilon^0_{\knu}-\epsilon^0_{\vcc{k}'\nu'}}}~\psi^0_{\vcc{k}'\nu'}(\rr) \nonumber
\end{eqnarray}
where
\begin{eqnarray}
    \mathcal{O}_\mathrm{xc}(\q) = \boldsymbol{\sigma}\cdot \left( \vcc{B}_\mathrm{xc}(\q,\rr) - \vcc{B}_\mathrm{xc}^{0} \right) \label{eq:model:Orot}
\end{eqnarray}
denotes the change of the magnetic exchange-correlation field relative to the ferromagnetic state magnetized along direction $\eFM$.  $\mathcal{O}_\mathrm{xc}$ is parallel to the local magnetization, and we assume that it rotates continuously in real-space and is of constant magnitude,
\begin{equation}
    \vcc{B}_\mathrm{xc} = B_\mathrm{xc}\, \underline{\underline{\mathcal{R}}}^{\hat{z} \rightarrow \erot}_{\hat{x} \rightarrow \eFM}\,\begin{pmatrix} \cos(\q \cdot \rr) \\ \sin(\q \cdot \rr) \\ 0 \end{pmatrix}\, ,
\end{equation}
where $\underline{\underline{\mathcal{R}}}$ is a rotation matrix that turns the local $z$-axis to $\erot$ and the local $x$-axis to the direction of the ferromagnetic state, $\eFM$. For our geometry as specified in \eqref{eq:model:DMI_q_derivative:main}, we have two choices for $\eFM$ (along the $z$ or $x$ axis, in the following called \emph{gauge I} and \emph{gauge II}, respectively) and obtain
\begin{eqnarray}\!\!
    \mathcal{O}_\mathrm{xc}^\mathrm{(I)}(q) \!&=&\! B_\mathrm{xc}\! \left[\phantom{-}\sin(q\, x)\,\sigma_x  + \left(\cos(q\, x) - 1 \right)\,\sigma_z \right] , \label{eq:model:Orot_xy_I} \\
    \mathcal{O}_\mathrm{xc}^\mathrm{(II)}(q) \!&=&\! B_\mathrm{xc}\! \left[ -\sin(q\, x)\,\sigma_z + \left(\cos(q\, x) - 1 \right)\,\sigma_x \right] . \label{eq:model:Orot_xy_II}
\end{eqnarray}
The wave-vector derivative in \eqref{eq:model:DMI_q_derivative:main} leads to a derivative of the expectation value in \eqref{eq:model:EDM}

\begin{equation}
    \frac{\partial}{\partial q} \braketmat{\psi_{\knu}}{\Hso}{\psi_\knu} =  \left\langle{\frac{\partial \psi_{\knu}}{\partial q}}\left\vert{\Hso}\right\vert{\psi_\knu}\right\rangle + \mathrm{c.c.}
\end{equation}
and the wave function, respectively, which is evaluated using \eqref{eq:model:psi-1st-order}

\begin{equation}
    \frac{\partial \psi_{\knu}(q,\rr)}{\partial q} =  \sum_{\vcc{k}'\nu' (\neq \vcc{k}\nu)}{ \frac{\braketmat{\psi^0_{\vcc{k}'\nu'}}{\partial \mathcal{O}_\mathrm{xc} / \partial q}{\psi^0_\knu}}{\epsilon^0_{\knu}-\epsilon^0_{\vcc{k}'\nu'}}}\,\psi^0_{\vcc{k}'\nu'}(\rr) \label{eq:model:psi-derivative}
\end{equation}
with
\begin{equation}
    \left. \frac{\partial \mathcal{O}_\mathrm{xc}^\mathrm{(I)}}{\partial q} \right\rvert_{q=0} = B_\mathrm{xc}\,\sigma_x\, x
        \quad \text{and} \quad
    \left. \frac{\partial \mathcal{O}_\mathrm{xc}^\mathrm{(II)}}{\partial q} \right\rvert_{q=0} = -B_\mathrm{xc}\,\sigma_z\,x\, , \label{eq:model:Orot_deriv_I_II}
\end{equation}
employing (\ref{eq:model:Orot_xy_I},  \ref{eq:model:Orot_xy_II}). For an arbitrary direction of $\eFM$, \eqref{eq:model:Orot_deriv_I_II} can be written in terms of the torque operator $\boldsymbol{\mathcal{T}} = -\boldsymbol{\sigma} \times \vcc{B}_\mathrm{xc}^{0}$,
\begin{equation}
    \left. \frac{\partial \mathcal{O}_\mathrm{xc}}{\partial q} \right\rvert_{q=0} = \mathcal{T}_y \, x,
\end{equation}
which has been termed the DMI operator in Ref.~\cite{Freimuth:arxiv:2018}.
Inserting everything leads to
\begin{eqnarray}
    D_\mathrm{s} &=& \Volfac \sum_{\knu}^\mathrm{occ.} \sum_{\nu'}^\mathrm{all} \frac{\braketmat{\psi^0_{\knu}}{\Hso}{\psi^0_{\knu'}} \braketmat{\psi^0_{\knu'}}{B_\mathrm{xc} \, \sigma_{x} ~x}{\psi^0_\knu} }{\epsilon^0_{\knu} - \epsilon^0_{\knu'}}\nonumber \\
    && +\,  \text{c.c.},\label{eq:model:Ds:perturb1}
\end{eqnarray}
where we restricted the analysis to gauge I. The states $\nu'$ run over occupied and unoccupied states. The sum over $\vcc{k}'$ drops out due to $\braketmat{\psi^0_{\knu}}{\Hso}{\psi^0_{\vcc{k}'\nu'}} \propto \delta_{\vcc{k},\vcc{k}'}$~\cite{Heide:Physica:2009}. The nominator is a product of spin-orbit and spin-torque-moment matrix elements and thus it is interpreted that spin-orbit matrix elements are weighted by the spin-torque strength. Let us recall that the states $\psi^0$ are eigenstates of the ferromagnet without SOC; hence they are of pure spin-character and eigenstates of $\sigma_z$. The Pauli-matrix $\sigma_x$ in the second bra-ket selects states $\nu'$ that have a different spin-character as compared to state $\nu$; \ie it selects spin-flip contributions.

To analyze \eqref{eq:model:Ds:perturb1} further, we break up the state index $\nu = (n, \sigma)$ into a band index $n$ and spin-index $\sigma \in \{\uparrow, \downarrow \}$ and express the wave function $ \psi^0$ in terms of spinor components:
\begin{eqnarray}
        \psi^0_{\kn \uparrow } = \begin{pmatrix} \mathring{\psi}^{\uparrow}_{\kn} \\ 0 \end{pmatrix}~, \qquad  \psi^0_{\kn \downarrow } = \begin{pmatrix} 0 \\ \mathring{\psi}^{\downarrow}_{\kn} \end{pmatrix}~ \label{eq:model:purespin:def}
\end{eqnarray}
Assuming that $B_\mathrm{xc}$ is approximately constant across the unit cell of the system, \eqref{eq:model:Ds:perturb1} is transformed to
\begin{eqnarray}
    D_\mathrm{s} &=& \VolfacBxc \sum_{\kn \sigma}^\mathrm{occ.} \sum_{n'}^\mathrm{all} \frac{\braketmat{\mathring{\psi}^\sigma_{\kn}}{\mathcal{H}_\mathrm{so}^{\sigma \sigma'}}{\mathring{\psi}^{\sigma'}_{\kn'}} \braketmat{\mathring{\psi}^{\sigma'}_{\kn'}}{x}{\mathring{\psi}^{\sigma}_\kn} }{\epsilon^0_{\kn \sigma} - \epsilon^0_{\kn' \sigma'}} \nonumber \\
    &&+\, \text{c.c.},  ~\qquad (\text{with}~\sigma'\neq \sigma) \label{eq:model:Ds:perturb2},
\end{eqnarray}
which establishes a connection between the interfacial DMI constant and the transition dipole moment on the right-hand side.

We quickly discuss gauge II: In this case, the second bra-ket in Eq.~\eqref{eq:model:Ds:perturb1} reads $\braketmat{\psi^0_{\knu'}}{\sigma_{z} ~x}{\psi^0_\knu}$ and now spin-conserving terms seem to play a role. However, now the ferromagnetic state is aligned along $x$ and corresponding $\psi^0_\knu$ are eigenstates of $\sigma_x$, so that $\sigma_z$ actually represents the spin-flip contributions with respect to the eigenstates of $\sigma_x$ in gauge II. Overall, we see that in both gauges, the second bra-ket selects spin-flip contributions with respect to the unperturbed eigenstates, and we may restrict the following analysis to gauge I.

\section{Approximations to the spin-torque moment and relation between DMI and $T_z$}
\label{sec:appendix:Bxcassumptions}

In advancing from (3) in the Sec.~\ref{sec:theoretical_derivation} [and from \eqref{eq:model:Ds:perturb1} to \eqref{eq:model:Ds:perturb2} in Appendix \ref{sec:appendix:Modelderivation}], we made the approximation of a uniform exchange-correlation field, $B_\mathrm{xc}$, across the unit cell. In the following we lift this model assumption. The second bra-ket of (3) and \eqref{eq:model:Ds:perturb1}, $\braketmat{\psi^0_{\knu'}}{\mathcal{T}_y  x}{\psi^0_\knu} $, turns into
\begin{equation}
    \braketmat{\psi^0_{\knu'}}{B_\mathrm{xc}(\vcc{r}) \, \sigma_{x} ~x}{\psi^0_\knu} \longrightarrow \braketmat{\varphi_2}{B_\mathrm{xc}(\vcc{r}) ~ x}{\varphi_1}\, , \label{eq:notes:braket2}
\end{equation}
with wavefunctions $\varphi_{1}$ and $\varphi_{2}$ defined in the main text. To evaluate these integrals we assume a tessellation of the solid in terms of muffin-tin spheres around atoms in which we expand the wavefunctions and the exchange field in spherical harmonics, $Y_{L}(\vcn{r})$, of unit vector $\vcn{r}$ and angular moment $L=(\ell,m)$,
\begin{eqnarray}
    \varphi_{1(2)}(\vcc{r})       &=& \sum_{L} \varphi^{(1(2))}_{L}(r) ~ Y_{L}(\vcn{r}) \\
    B_\mathrm{xc}(\vcc{r}) &=& \sum_{L} B^\mathrm{xc}_{L}(r)       ~ Y_{L}(\vcn{r})\, . \label{eq:notes:Bxc-expansion}
\end{eqnarray}
The spherical harmonics are orthonormal, $\langle Y_L\vert Y_{L'} \rangle = \delta_{LL'}$.

We first assume $B_\mathrm{xc}(\vcc{r})$ to be spherically symmetric; \ie only the term with $L=0$ in \eqref{eq:notes:Bxc-expansion} remains. Substituting $B_\mathrm{xc}(\vcc{r})$ in \eqref{eq:notes:braket2} and separating radial and angular integrals yields
\begin{eqnarray}
    \braketmat{\varphi_2}{B^\mathrm{xc}_{0}(r)\, x}{\varphi_1} &=& \frac{1}{\sqrt{4\pi}}\sum_{L_1,L_2} \braketmatsmall{\varphi^{(2)}_{L_2}}{B^\mathrm{xc}_{0}\,r}{\varphi^{(1)}_{L_1}}_{|_R} \nonumber \\
    && \times \braketmat{L_2}{\hat{x}}{L_1}
\end{eqnarray}
with
\begin{equation}
   \braketmat{\alpha}{f}{\gamma}_{|_R}  := \int r^2 \td{r}\,{\alpha^*(r)\, f(r)\, \gamma(r)}
\end{equation}
and $\hat{x}=x/r$. The angular part is of the same form as in the main text and hence the same transitions (\cf Table~1 of the main text) constitute the finite DMI. For the radial part we consider the example of the main text taking the radial representation of the state $\ket{\varphi_1}$, $\braket{r}{\varphi_1} = \alpha(r) \ket{s} + \beta(r) \ket{p_z}$, and state $\ket{\varphi_2}$, $\braket{r}{\varphi_2} = \gamma(r) \ket{p_x}$, with proper normalization $\int r^2\td{r} \, (\abs{\alpha}^2 + \abs{\beta}^2)=1$ and  $\int r^2\td{r} \, \abs{\gamma}^2=1$, which transforms the prefactor $B_\mathrm{xc} ~\alpha \beta^*$ from the main text into
\begin{eqnarray}
    \longrightarrow \braketmat{\beta}{V_\mathrm{so}}{\gamma}_{|_R}  \braketmat{\gamma}{ B^\mathrm{xc}_{0}\,r}{\alpha}_{|_R}
\end{eqnarray}
where the spin-orbit operator has been rewritten as $\Hso = V_\mathrm{so}(r) ~ \vcc{L}\cdot \vcc{S}$.

Going beyond the spherical approximation, the next non-spherical term to consider is of order $\ell=1$. Due to the uniaxial symmetry of the MMLs the terms $m\pm 1$ vanish and the remaining contribution reads $B_{(1,0)}(r)Y_{(1,0)}(\vcn{r}) \propto B_{(1,0)}(r)\hat{z}$, which yields for \eqref{eq:notes:braket2}
\begin{equation}
    \braketmat{\varphi_2}{B^\mathrm{xc}_{(1,0)}(r) \hat{z} x}{\varphi_1} = \braketmat{\gamma}{B^\mathrm{xc}_{(1,0)} \, r}{\alpha}_{|_R} \braketmat{L_2}{\hat{z}\hat{x}}{L_1}
\end{equation}
The operator acting on the angular part is of angular momentum $\ell=2$, \ie $\hat{z}\hat{x} \sim Y_{2,-1} - Y_{2,1}$. Hence, the determination of possible transitions is governed by the symmetry of the Gaunt coefficients $G_{\ell_1 \, 2 \, \ell_2}^{m_1 \, \pm1 \, m_2}$, and only $s-d$, $p-p$ and $d-d$-transitions remain (neglecting transitions to $f$ states and beyond) with the additional selection rule $m_1 = m_2 \pm 1$.

On the other hand, we analyze the magnetic dipole moment,
\begin{eqnarray}
    \!\!\!\!\!\!\!\!\!\!\!\!\!\!\!\!\!T_z &=& \frac{\hbar}{2}\sum_\knu \braketmat{\psi_\knu}{ Q_{zx}\sigma_x + Q_{zy}\sigma_y + Q_{zz}\sigma_z}{\psi_\knu} \label{eq:appendix:Tz:Qsigma_explicit}\\
    &\propto& \sum_\knu \braketmat{\psi_\knu}{ \hat{z}\hat{x}\,\sigma_x + \hat{z}\hat{y}\,\sigma_y + (\hat{z}^2-\frac{1}{3})\,\sigma_z}{\psi_\knu} \label{eq:appendix:Tz:xyz} \\
    &\propto& \sum_\knu \braketmat{\psi_\knu}{ Y_{xz}\sigma_x + Y_{yz}\sigma_y + \frac{2}{\sqrt{3}} Y_{z^2}\sigma_z}{\psi_\knu}\!, \label{eq:appendix:Tz:dxyz}
\end{eqnarray}
where $Y_{xz}$, etc., denote real spherical harmonics in Cartesian coordinates. Interestingly, the first term in \eqref{eq:appendix:Tz:xyz} is exactly the same operator as the one that appears in \eqref{eq:notes:braket2} if the non-spherical correction to the exchange field is considered ($B_\mathrm{xc}(\vcc{r}) \rightarrow B^\mathrm{xc}_{(1,0)} \hat{z}$). However, if the magnetic system assumes the ferromagnetic state, and the SOC is neglected, the state $\ket{\psi_\nu}$ is an eigenstate of $\sigma_z$, the $T_z$ reduces to the third term in \eqref{eq:appendix:Tz:xyz} proportional to $Q_{zz}$ and the spin-flip terms $\propto \sigma_x, \sigma_y$, or $Q_{xz}$ and $Q_{yz}$, respectively, which are the terms that contribute to the DMI, disappear. This underlines that it is the spin-orbit contribution to the wave functions that activates the spin-flip contributions which relate $T_z$ and the DMI.

\section{Calculation of $\vcc{p}^\text{el}$ and $\vcc{T}$}
\label{sec:appendix:dipolesDefinition}

In the FLAPW method, the charge density $\rho(\vcc{r})$ and vector-spin density $\vcc{s}(\vcc{r})$ within the muffin-tin (MT) spheres around each atom are naturally available in terms of spherical harmonics expansions as
\begin{eqnarray}
    \rho(\vcc{r}) &=& \sum_{L} \rho_{L}(r) Y_{L}(\hat{\vcc{r}}), \label{elecDipoleExp} \\ 
    \vcc{s}(\vcc{r}) &=& \sum_{L} \vcc{s}_{L}(r) Y_{L}(\hat{\vcc{r}}),
\end{eqnarray}
where $\vcc{r}$ is the position vector relative to the atomic nucleus. These quantities are the starting point for the calculation of the electric dipole moment $\vcc{p}^\text{el}$ and of the intraatomic magnetic dipole moment $\vcc{T}$, which are defined in Eqs.~(4) and~(5). A more practical formulation of (5) for a numerical evaluation within the FLAPW method is based on the vector-spin density $\mathbf{s}(\mathbf{r})$ and reads
\begin{eqnarray}
    \mathbf{T} &=& \int_{\Omega_\mathrm{MT}}{ \left[\vcc{s}(\vcc{r}) - 3 \vcn{r} (\vcn{r} \cdot \vcc{s}(\vcc{r})) \right]}\, \mathrm{d}\mathbf{r}\, ,\label{magneticDipoleDefEq}
\end{eqnarray}
where $\vcn{r} = \mathbf{r}/\vert \mathbf{r} \vert$ is the normalized position vector~\cite{Oguchi:PRB:2004}.

By expressing the unit vector in terms of spherical harmonics as
\begin{equation}
\hat{\vcc{r}} = \sqrt{\frac{2\pi}{3}}\left( \begin{matrix} Y_{1,-1}(\hat{\vcc{r}}) - Y_{1,1}(\hat{\vcc{r}}) \\ i Y_{1,-1}(\hat{\vcc{r}}) + i Y_{1,1}(\hat{\vcc{r}}) \\ \sqrt{2} Y_{1,0}(\hat{\vcc{r}}) \end{matrix} \right) =: \sum_{m=-1}^{1} \vcc{g}_{m} Y_{1,m}(\hat{\vcc{r}}) \label{eq:appendix:unit_r}
\end{equation}
and making use of the spherical harmonics expansion~\eqref{elecDipoleExp} the electric dipole moment can now be written as
\begin{eqnarray}
    \vcc{p}^\text{el} &=& \int_{\Omega_\text{MT}} \sum_{\ell,m} \sum_{m'=-1}^1\!\! \rho_{\ell,m}(r)\, r \, \vcc{g}_{m'} Y_{1,m'}(\hat{\vcc{r}}) Y_{\ell,m}(\hat{\vcc{r}})\, \text{d}^3r \notag \\
    &=& \sum_{m=-1}^1 (-1)^m \vcc{g}_{m} \int_{0}^{R_\mathrm{MT}}  \!\! \rho_{1,-m}(r)\, r^3 \,\text{d}r\, .
\label{elecDipoleCalcEq}
\end{eqnarray}
The latter simplification was achieved by angular integration over the spherical harmonics, yielding $\delta_{1,\ell}~\delta_{m,m'}$,
where $\Omega_\text{MT}$ and $R_\text{MT}$ are volume and radius of the MT sphere centered at the respective atom, respectively.

Similarly the intra-atomic magnetic dipole moment can be written as
\begin{eqnarray}
    \vcc{T} &=& \sqrt{4\pi} \int_{0}^{R_\mathrm{MT}} \left(\vcc{s}_{0,0}(r) - 3 \vcc{A}_{0,0}(r)\right) \, r^2 \, \text{d}r
    \label{magDipoleCalcEq}
\end{eqnarray}
where $\vcc{A}_{0,0}(r) = \sum_{\ell,m} \sum_{m'=-1}^{1} \vcc{g}_{m'} B_{\ell,m}(r) G_{0\ell1}^{0mm'}$, $B_{\ell,m}(r) = \sum_{\ell',m'} \sum_{m''=-1}^{1} \vcc{g}_{m''} \vcc{s}_{\ell',m'}(r) G_{\ell \ell '1}^{mm'm''}$, and $G_{\ell \ell' \ell''}^{mm'm''}$ are the Gaunt coefficients.

It should be noted that the $l=1$ character of the unit vector, together with the rules for non-vanishing Gaunt coefficients and cancellations in the equations, leads to very few $(\ell,m)$ combinations in the expansion of the charge and vector-spin densities that are relevant for the calculation of $\vcc{p}^\text{el}$ and $\vcc{T}$. For $\vcc{p}^\text{el}$ only the $\ell=1$ components of the charge density are relevant and for $\vcc{T}$ only the $\ell=2$ components of the vector-spin density.

\end{document}